\documentclass[12pt]{article}

\usepackage[T1]{fontenc}
\usepackage{newtxtext,newtxmath}

\usepackage{graphicx}
\usepackage{subfig}

\usepackage[letterpaper,margin=1in]{geometry}

\renewenvironment{abstract}
	{\quotation}
	{\endquotation}

\date{}

\makeatletter
\renewcommand{\fnum@figure}{\textbf{Figure \thefigure}}
\renewcommand{\fnum@table}{\textbf{Table \thetable}}
\makeatother

\usepackage{scicite}

\usepackage{url}

\usepackage{physics}

\def\scititle{
	Order-of-magnitude extension of qubit lifetimes with a decoherence-free subspace quantum error correction code
}
\title{\bfseries \boldmath \scititle}

\author{
	Shival~Dasu$^{1\ast}$,
        Ben ~Criger$^{2}$,
        Cameron ~Foltz$^{1}$,
        Justin A. ~Gerber$^{1}$,\and
        Christopher N. ~Gilbreth$^{1}$,
        Kevin~Gilmore$^{1}$,
        Craig~A.~Holliman$^{3}$,
        Nathan~K.~Lysne$^{3}$,\and
        Alistair.~R.~Milne$^{4}$,
        Daichi ~Okuno$^{3}$,
        Grahame~Vittorini$^{1}$,
        David~Hayes$^{1}$\and
	\small$^{1}$ Quantinuum, 303 South Technology Ct., Broomfield, CO 80021, USA\and
        \small$^{2}$ Quantinuum, Terrington House, 13–15 Hills Road, Cambridge, CB2 1NL, UK\and
	\small$^{3}$ Quantinuum K.K., Otemachi Financial City Grand Cube 3F, 1-9-2 Otemachi, Chiyoda-ku, Tokyo, Japan\and
        \small$^{4}$ Quantinuum, Partnership House, Carlisle Place, London SW1P 1BX, UK. \and
	\small$^\ast$Corresponding author. Email: shival.dasu@quantinuum.com
}

\begin{document}

\maketitle

\begin{abstract} \bfseries \boldmath

Constructing an efficient and robust quantum memory is central to the challenge of engineering feasible quantum computer architectures. Quantum error correction codes can solve this problem in theory, but without careful design it can introduce daunting requirements that call for machines many orders of magnitude larger than what is available today. Bringing these requirements down can often be achieved by tailoring the codes to mitigate the specific forms of noise known to be present. Using a Quantinuum H1 quantum computer, we report on a robust quantum memory design using a concatenated code, with the low-level code designed to mitigate the dominant source of memory error, and a higher-level error correction scheme to enable robust computation. The resulting encoding scheme, known as a decoherence-free subspace quantum error correction code, is characterized for long probe times, and shown to extend the memory time by over an order of magnitude compared to physical qubits.
\end{abstract}

\noindent
The quantum charge-coupled device (QCCD) trapped-ion quantum computer architecture introduced in Ref.~\cite{Wineland98} performs computations by shuttling ions to specialized regions for different operations, such as storage, gating, or measurement. To protect the quantum information from decoherence while in transit, Ref.~\cite{Kielpinski2002} proposed encoding qubits in a decoherence-free subspace (DFS)~\cite{Lidar1998} with ion-qubits paired to form a simple logical qubit. The logical encoding would pair physical qubits in an opposite fashion ($\ket{0_L}=\ket{\uparrow\downarrow}$, $\ket{1_L}=\ket{\downarrow\uparrow}$) and be transported together. This scheme ensures that when magnetic fields fluctuate in time or space and individual physical qubits acquire an uncontrolled phase shift, $\ket{\downarrow}\rightarrow e^{i\phi}\ket{\downarrow},\ket{\uparrow}\rightarrow e^{-i\phi}\ket{\uparrow}$, the logical qubits remain stable, $\ket{\uparrow\downarrow}\rightarrow \ket{\uparrow\downarrow}$, a property which has been demonstrated in several experiments~\cite{Kielpinski2001,Monz2009,Quiroz2024}. This DFS is also a stabilizer code with a single $-ZZ$ stabilizer and parameters $[[2,1,1]]$, meaning two physical qubits encode a single logical qubit with distance $d=1$, (bit flip errors are detectable in the DFS code, but phase errors are not). While this encoding may mitigate predominant sources of memory error in trapped ions, additional error mechanisms are also present, making quantum error correction (QEC) a necessity for large-scale computations. To this end, the community has recently made impressive progress in demonstrating general QEC to protect against general noise environments~\cite{RyanAnderson2021,Sundaresan2023,Ni2023,Sivak2023,Postler2024,Lachance2024,Bluvstein2024,Paetznick2024,RyanAnderson2024,Acharya2024}. These general QEC demonstrations are truly important milestones in the march toward fault-tolerance, but practical realizations of large-scale quantum computers may greatly benefit from QEC codes tailored for specific noise environments, and memory errors have been significant limitations in early QCCD demonstrations of QEC~\cite{RyanAnderson2021,Paetznick2024}. Theoretical proposals for tailored codes to mitigate structured noise environments include asymmetric distance~\cite{Tuckett2019}, Clifford deformation~\cite{Dua2024}, or novel decoding strategies~\cite{Higgott2023}. 
Another approach is to build in passive stability at a low level of concatenation, where a logical encoding is nested within another encoding, similar to the approach of dual-rail qubits~\cite{Levine2024,Koottandavida2024}. If a logical qubit were passively stable to memory noise, it would not require error correction while in memory, significantly easing resource requirements for circuits where logical qubits are frequently idling.

In this work, we demonstrate the concatenation of a passively stable $[[2,1,1]]$ DFS code with a $[[5,1,3]]$ QEC code~\cite{Laflamme96} to form a $[[10,1,4]]$ DFS-QEC code~\cite{LidarDFSQEC1999,Boulant2005,Hu2021,Ouyang2021,Dash2024}. Using a Quantinuum H1 quantum computer, we benchmark the quantum memory of three different qubits in similar noise environments: physical qubits, DFS qubits, and DFS-QEC qubits. In the case of the DFS-QEC qubits, we report the fidelities using two different modes, both of which are relevant to the study of fault-tolerant quantum computer architectures: the first being a pure QEC mode, and the second a quantum error detection (QED) mode to post-select some weight-larger-than-one errors while still correcting all weight-one errors. The first mode is relevant for benchmarking how well a quantum computer can store quantum information in memory, whereas the second mode is relevant to the storage of resource states such as logical ancilla or magic states. The memory quality is quantified by measuring the fidelity as a function of time for different initial states. We report the raw measurements and further quantify decay rates using simple models fit to the data. Our results show over an order-of-magnitude improvement in the lifetime of the encoded qubit over the physical layer qubit, a significant improvement over previous logical memory break-even demonstrations~\cite{Acharya2024,Ni2023,Sivak2023,Lachance2024}.

\subsection*{Experiments}
The H1 system uses approximate clock-state qubits~\cite{Olmschenk2007} in the $^{2}S_{1/2}$ hyperfine manifold of $^{171}\text{Yb}^+$. A clock-state qubit provides a good starting point for a quantum memory due to the reduced sensitivity of the qubit frequency to magnetic field fluctuations~\cite{Langer05,Wang2021}. We operate at a bias field of 3.27 G, resulting in a first-order qubit frequency sensitivity of $2032 \times \delta B$ Hz/G where $\delta B$ is the magnetic field noise amplitude. We use passive shielding to help minimize any noise and periodically measure the qubit frequency at many positions along the trap to account for fluctuations in DC magnetic fields or shifts due to AC magnetic fields generated by currents in the trap. As qubits are transported to different zones, the varying frequency leads to differential phases which are tracked with a software routine and compensated for by adjusting the laser phases appropriately during gating~\cite{RyanAnderson2021}.

\begin{figure}[ht]
\includegraphics[width=1.0\textwidth]{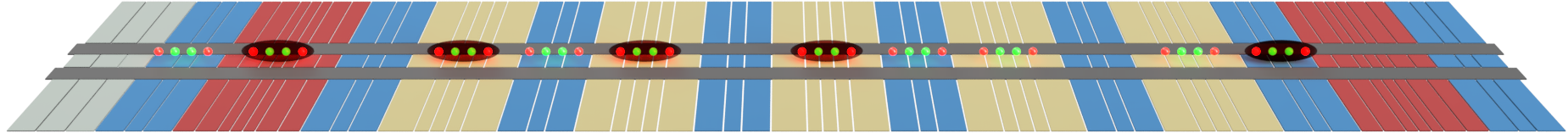}
\caption{\textbf{An illustration of the H1 trap~\cite{Pino2020}.} The trap is configured with 20 $^{171}\text{Yb}^+$ qubit ions (red circles) and 20 $^{138}\text{Ba}^+$ coolant ions (green circles), and has four different types of zones: a loading zone on the left (grey), eight small storage zones (blue), two large storage zones (red), and five gating zones (yellow). In this configuration, three DFS pairs reside in gate zones, and two DFS pairs reside in storage zones.}
\label{fig:trap}
\end{figure}

To make a fair comparison, all of our experiments use ten data qubits, the minimum for the DFS-QEC code experiment. The H1 system has five gating zones (Fig.~\ref{fig:trap}), each of which can store two qubits and two $^{138}\text{Ba}^+$ ions used for sympathetic cooling~\cite{Pino2020}. Our memory experiments rely on letting qubits idle for extended periods of time, and co-locating the DFS pairs minimizes differential phase accumulation due to spatial inhomogeneities in the magnetic-field. After we have paired the qubits together in the physical, DFS, and DFS-QEC experiments, we then allow the compiler to store them in a typical configuration during sleep, which either has two or three of the pairs idling in gate zones and the rest in storage zones. This ensures that we are exposing the qubits to noise that is typical across the entire trap. In the cases of the physical and DFS-QEC qubits, we characterize the memory using an informationally complete set of states: the six different eigenstates of the single-qubit Pauli operators, $\ket{\psi} \in \{\ket{0}, \ket{1}, \ket{+}, \ket{-}, \ket{+i}, \ket{-i}\}$. Due to machine-time constraints, we characterized the DFS qubits using a smaller set of initial states, $\{\ket{1},\ket{+},\ket{+i}\}$ and assume similar performance for the orthogonal set of states.

In the physical layer memory experiments, we initialize the state, wait for increments of $\sim2$~s, then measure in the corresponding initialization basis and record the probability $p_\psi(t)$ that the state is unchanged. The $[[2,1,1]]$ DFS code experiments were analogous to the physical layer experiments; we probed five copies of the DFS code in each shot, one in each gating zone, and averaged over the five gate zones. Using the smaller set of states mentioned earlier, we initialize, wait for increments of $\sim2$~s, then measure in the corresponding basis.

In the DFS-QEC code experiments, we begin by creating the logical Pauli-basis states for the $[[10,1,4]]$ DFS-QEC code (circuits shown in Fig.~\ref{sup fig:init circuits}). The state preparation circuits are not fault-tolerant, but this only determines the initial fidelity and does not affect memory performance over time. Once the states are initialized, they idle, and then undergo a fault-tolerant QEC cycle, completing a cycle approximately every 2.89 s for up to $\sim23$~s. Ideally, the DFS pairs would stay co-located for the entire circuit, but re-sorting is necessary during syndrome extraction which forces the DFS pairs to be temporarily separated. After syndrome extraction, we call for a zero-angle entangling gate between the DFS pairs, bringing the pairs back to being co-located, then a sleep command is issued to begin the next idling period. The syndrome extraction circuits are a modification of Reichardt's parallel flagging scheme ~\cite{Reichardt2018} for the five-qubit code to the DFS case (Figs.~\ref{sup fig:fault-tolerant SE 1} and \ref{sup fig:fault-tolerant SE 2}). If any flags or syndromes are non-trivial we also extract syndromes using the unflagged circuits in Fig.~ \ref{sup fig:unflaggedSE1}  and \ref{sup fig:unflaggedSE2}. The DFS syndromes are extracted via the fault-tolerant circuit in Fig.~\ref{sup fig:fault-tolerant SE DFS}. Our decoder was implemented in Rust compiled to Wasm~\cite{RyanAnderson2022} and executed in real time with active corrections applied at each round of syndrome extraction. At the end of the experiment, we measure the qubits in the appropriate basis and record the probability of being in the correct state. For a more detailed description of this procedure, our decoder, and a full analysis of the fault tolerance of our QEC cycle, see the Supplementary Materials. 

After recording individual state fidelities, we evaluate multiple figures of merit. Using the individual fidelities, we calculate the average state fidelity of the memory channel, $F_a = \frac{1}{6} \sum_\psi p_\psi$, and the process fidelity $F_p = ((d+1) F_a - 1)/d$~\cite{Nielsen2002}, where the dimension $d=2$ for a single qubit. As suggested in ~\cite{Bermudez2017,Xu2018}, we also assess a quantity called the \textit{integrity} $\mathcal{R}(\Phi)$ which is a function of the memory channel $\Phi$ and describes the best possible guarantee on how well quantum information is preserved by the channel. The integrity is computed by taking the minimum state fidelity of the different bases, $p_{\text{worst}}=\text{min}_{\psi}\frac{1}{2}(p_{\psi}+p_{\psi_{\perp}})$, and computing $\mathcal{R}=2|p_{\text{worst}}-\frac{1}{2}|$. This expression differs from the original expression in Ref.~\cite{Xu2018} by including the absolute value to account for fidelities $<0.5$ caused by coherent errors that could in principle be corrected.

\subsection*{Results}

\begin{figure}[ht]
 \centering
   \includegraphics[width=1.0
  \linewidth]{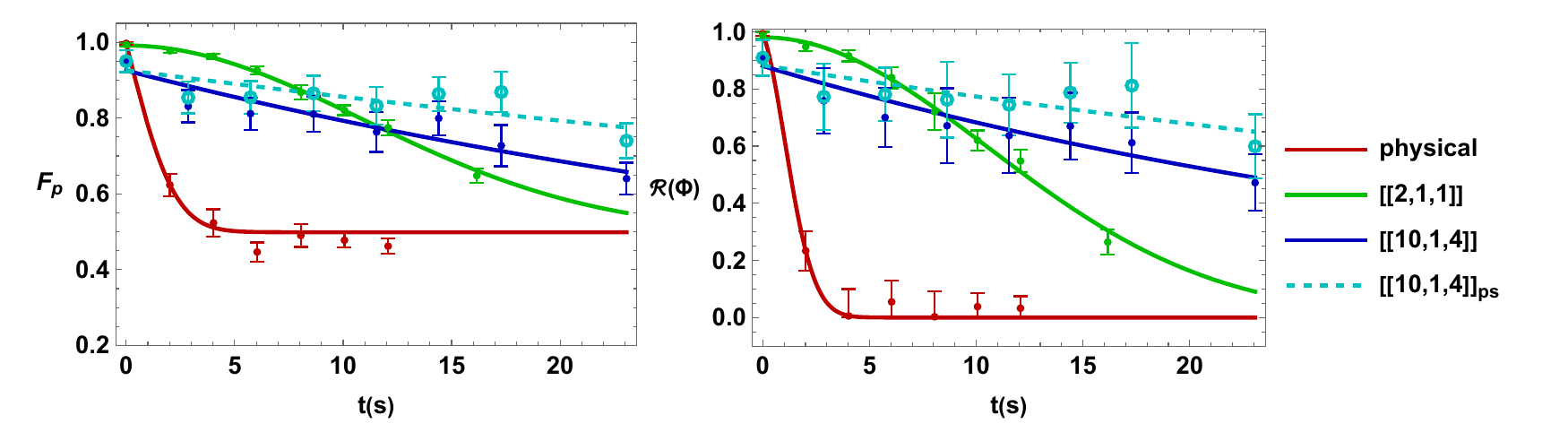}
\caption{\textbf{The process fidelity $F_p$ and integrity measure $\mathcal{R}(\Phi)$ of the three different qubits investigated in this work.} We use an informationally complete set of states for the physical and DFS-QEC $[[10,1,4]]$ encodings, but use an incomplete set of measurements for the DFS $[[2,1,1]]$ qubits, measuring the fidelity of $\{\ket{1},\ket{+},\ket{+i}\}$ and assuming the orthogonal states have approximately the same fidelity as the orthogonal states. We also show the results when post-selection is used in the DFS-QEC code, denoted $[[10,1,4]]_{\text{ps}}$, and show the post-selection rates in Fig.~\ref{fig:RetentionRates}.}
\label{fig:Integrity}
\end{figure}

As seen in Fig.~\ref{fig:Integrity}, the process fidelity and integrity of the DFS and DFS-QEC qubits are greatly enhanced over the physical qubit. To help quantify the advantage of these encodings, we fit the data to decay models whose derivations are discussed in the Supplementary Materials. The models serve as a guide to the eye in Fig.~\ref{fig:Integrity} and allow us to separate the state preparation and measurement (SPAM) error from the memory error. The physical and DFS qubit decay models have both Gaussian and exponential decay terms to capture coherent and incoherent dephasing, but we find that the individual states are accurately modeled by \textit{either} a Gaussian \textit{or} an exponential. 

For the physical and DFS qubits, the superposition states $\{\ket{+},\ket{-},\ket{+i},\ket{-i}\}$ are dominated by Gaussian decay as their errors are largely coherent in individual zones but inhomogenous over different regions in the trap. Upon averaging over the different regions, the coherent errors result in a Gaussian decay of the superposition states with a time constant of $\sim0.57$ Hz. The DFS code outperforms the physical layer memory by an order of magnitude, with a time constant of $\sim0.065$ Hz for the superposition states.

As shown in the Supplementary Material, the individual physical layer states show different sensitivities. The experiments using the computational basis states $\ket{0}$ and $\ket{1}$ show slow decay rates compared to the superposition states $\{\ket{+},\ket{-},\ket{+i},\ket{-i}\}$. This asymmetry is typical in trapped-ion experiments where the dominant memory error process is caused by stochastic and coherent $\hat{Z}$ processes, and it implies that the physical and DFS qubits' integrity is dominated by the superposition states. We note that the measurement protocol used here~\cite{Olmschenk2007} does not distinguish between $\ket{1}$ and the Zeeman states $\ket{F=1,m_f=\pm1}$ which form a leakage subspace. Except for the $\ket{1}$ probe, all physical-level experiments are measured out in the $\ket{0}$ state. For example, the $\ket{+}$ state is probed by initializing to $\ket{0}$ via optical pumping~\cite{Olmschenk2007}, then is rotated to $\ket{+}$ via a single-qubit gate, allowed to idle, rotated back to $\ket{0}$, and measured. As the population in the leakage subspace will not return to $\ket{0}$ after the final rotation, superposition state fidelities are sensitive to these excess $\ket{1}$ outcomes. However, the relative magnitude of any leakage errors compared to dephasing errors is not quantifiable in these experiments, and we leave this to future work.

\begin{figure}[ht]
 \centering
  \includegraphics[width=.7
  \linewidth]{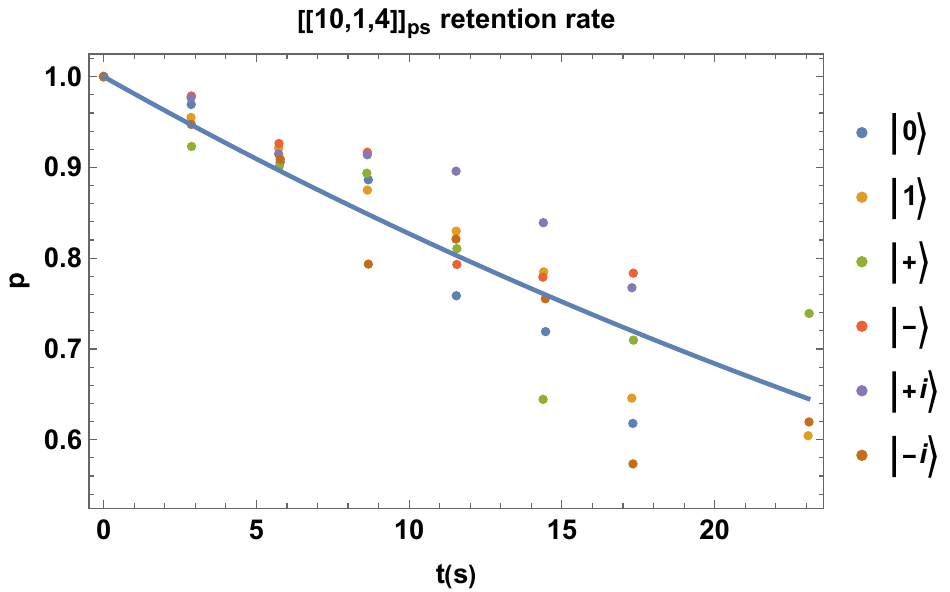}
\caption{\textbf{The retention rates for the $[[10,1,4]]_{\text{ps}}$ code using post-selection.} As shown by the legend, we plot the retention rates for the six different states probed, and we also plot a best fit decay curve of the average retention rate. The decay curve is of the form $p(t)=\text{exp}[-\eta t]$ and the fitting parameter is $\eta=0.019~\text{s}^{-1}$.}
\label{fig:RetentionRates}
\end{figure}

The [[10,1,4]] qubits behave very differently, and the data is well described by an exponential decay with a time constant of $\sim0.02$ Hz. When using post-selection, the time constant is improved to $\sim0.009$ Hz. As shown in Fig.~\ref{fig:RetentionRates}, the retention rate when using post-selection is approximately $\text{exp}[-\eta t]$, where $\eta=0.019~\text{s}^{-1}$ and each QEC cycle has a retention rate of $\sim95\%$. Unlike the physical and DFS qubits, the individual [[10,1,4]] states show more uniform decay rates. It is worth noting that the logical measurements are sensitive to leakage as well. If a single qubit leaks, this will affect the parity of the logical measurement if the measured codeword should have had a measurement corresponding to $\ket{0}$ for that qubit. This process is non-deterministic. For example, if an even-index qubit leaks, this will have a 50\% chance of flipping the measured parity of the logical Z operator, since the distribution of 0s and 1s in codewords is uniform over each qubit.

The different forms of the encoded and unencoded decay models make their direct comparison problematic, but one does not need the aid of the fitting models to see the advantage of the encodings over the physical layer memory. Despite the different decay forms, we can examine the $1/e$ lifetimes: for the process fidelity, we define it as the time to reach $F=1-1/2e$ (since $1/2$ is the asymptotic value), and for the integrity measure we define it as the time to reach $\mathcal{R}=1/e$. Using these definitions and the fit models, the $1/e$ lifetimes suggest the DFS-QEC code to physical memory advantage as measured by process fidelity and integrity is $\sim10$x and $\sim13$x respectively. Likewise, the post-selected DFS-QEC code extends the process fidelity and integrity lifetimes over the physical qubit lifetime by $\sim20$x and $\sim25$x. However, we caution that these improvement factors may not apply to the small-error regime. 

While the fitting models indicate DFS-QEC encoding to have the longest life-time as defined by $1/e$, the models also indicate the DFS encoding is superior at short times. However, this artifact is largely attributable to SPAM errors in the DFS-QEC qubit and the coarse-grained measurements, as the decay between QEC cycles should not be exponential~\cite{Xu2018}. Since the decay is driven by coherent error, which scales like $t^2$ at the physical level, the DFS-QEC code should decay like $t^4$ when only using QEC, and should decay like $t^6$ when also using post-selection. We leave an experimental confirmation of this to future work. All fitting parameters are shown in Table~\ref{Table:FittingParameters}.

\begin{table}
\resizebox{\textwidth}{!}{%
\begin{tabular}{|c|ccc|ccc|cc|cc|}
\hline

\multicolumn{1}{|c|}{} &
  \multicolumn{3}{c|}{physical} &
  \multicolumn{3}{c|}{[[2,1,1]]} &
  \multicolumn{2}{c|}{[[10,1,4]]} &
  \multicolumn{2}{c|}{$[[10,1,4]]_{\text{ps}}$} \\ \hline

\multicolumn{1}{|c|}{} &
  \multicolumn{1}{c|}{$\epsilon_{p,s}$} &
  \multicolumn{1}{c|}{$\gamma_p$ ($s^{-1}$)} &
  \multicolumn{1}{c|}{$2^{-1/2}\Gamma_p$ ($s^{-1}$)} &
  \multicolumn{1}{c|}{$\epsilon_{d,s}$} &
  \multicolumn{1}{c|}{$\gamma_d$ ($s^{-1}$)} &
  \multicolumn{1}{c|}{$2^{-1/2}\Gamma_d$ ($s^{-1}$)} &
  \multicolumn{1}{c|}{$\epsilon_{q,s}$} &
  \multicolumn{1}{c|}{$2\epsilon_{q,m}\tau^{-1}(s^{-1})$} &
  \multicolumn{1}{c|}{$\epsilon_{q,s}$} &
  \multicolumn{1}{c|}{$2\epsilon_{q,m}\tau^{-1}(s^{-1})$} \\ \hline\hline
$\ket{0}$  & \multicolumn{1}{c|}{1.3(8)e-3} & \multicolumn{1}{c|}{1(2)e-4} & \multicolumn{1}{c|}{0}         & \multicolumn{1}{c|}{N/A}        & \multicolumn{1}{c|}{N/A}      & \multicolumn{1}{c|}{N/A}       & \multicolumn{1}{c|}{4(2)e-2} & \multicolumn{1}{c|}{2.6(5)e-2} & \multicolumn{1}{c|}{5(2)e-2} & \multicolumn{1}{c|}{9(5)e-3} \\ \hline
$\ket{1}$  & \multicolumn{1}{c|}{5(2)e-3}   & \multicolumn{1}{c|}{0(5)e-4} & 0         & \multicolumn{1}{c|}{3(2)e-3}   & \multicolumn{1}{c|}{0(1)e-3} & \multicolumn{1}{c|}{0}         & \multicolumn{1}{c|}{5(2)e-2} & \multicolumn{1}{c|}{2.3(5)e-2} & \multicolumn{1}{c|}{5(2)e-2} & \multicolumn{1}{c|}{1.1(5)e-2} \\ \hline
$\ket{+}$   & \multicolumn{1}{c|}{2(5)e-3}   & \multicolumn{1}{c|}{0}       & \multicolumn{1}{c|}{5(1)e-1}   & \multicolumn{1}{c|}{1(3)e-3}   & \multicolumn{1}{c|}{0}       & \multicolumn{1}{c|}{6.7(2)e-2} & \multicolumn{1}{c|}{4(2)e-2} & \multicolumn{1}{c|}{2.1(5)e-2} & \multicolumn{1}{c|}{3(3)e-2} & \multicolumn{1}{c|}{1.1(6)e-2} \\ \hline
$\ket{-}$   & \multicolumn{1}{c|}{1(3)e-3}   & \multicolumn{1}{c|}{0}      & \multicolumn{1}{c|}{6(1)e-1}   & \multicolumn{1}{c|}{N/A}        & \multicolumn{1}{c|}{N/A}      & \multicolumn{1}{c|}{N/A}        & \multicolumn{1}{c|}{6(2)e-2} & \multicolumn{1}{c|}{9(5)e-3} & \multicolumn{1}{c|}{6(2)e-2} & \multicolumn{1}{c|}{3(4)e-3}  \\ \hline
$\ket{+i}$  & \multicolumn{1}{c|}{2(3)e-3}   & \multicolumn{1}{c|}{0}       & \multicolumn{1}{c|}{5.7(5)e-1} & \multicolumn{1}{c|}{1.1(3)e-2} & \multicolumn{1}{c|}{0}       & \multicolumn{1}{c|}{6.3(2)e-2} & \multicolumn{1}{c|}{3(2)e-2} & \multicolumn{1}{c|}{1.5(6)e-2} & \multicolumn{1}{c|}{4(2)e-2} & \multicolumn{1}{c|}{6(6)e-3}   \\ \hline
    $\ket{-i}$  & \multicolumn{1}{c|}{9(17)e-4}  & \multicolumn{1}{c|}{0}       & \multicolumn{1}{c|}{6.3(7)e-1} & \multicolumn{1}{c|}{N/A}        & \multicolumn{1}{c|}{N/A}      & \multicolumn{1}{c|}{N/A}        & \multicolumn{1}{c|}{3(2)e-2} & \multicolumn{1}{c|}{2.6(7)e-2} & \multicolumn{1}{c|}{3(2)e-2} & \multicolumn{1}{c|}{1.3(5)e-2} \\ \hline
mean  & \multicolumn{1}{c|}{2(2)e-3}  & \multicolumn{1}{c|}{N/A}       & \multicolumn{1}{c|}{N/A} & \multicolumn{1}{c|}{N/A}        & \multicolumn{1}{c|}{N/A}      & \multicolumn{1}{c|}{N/A}        & \multicolumn{1}{c|}{4(1)e-2} & \multicolumn{1}{c|}{2.0(7)e-2} & \multicolumn{1}{c|}{5(1)e-2} & \multicolumn{1}{c|}{9(4)e-3} \\ \hline
\end{tabular}%
}
\caption{\textbf{Fit parameters for decay models of individual states for the three different qubit definitions, including results of the DFS-QEC code using post-selection, denoted $[[10,1,4]]_{\text{ps}}$.} ~As detailed in the Supplementary Materials, the physical and DFS $[[2,1,1]]$ qubits' decay model
$F_i(t)=1/2+(1/2-\epsilon_{i,s})\text{exp}[-\gamma_it-\frac{1}{2}(\Gamma_it)^2]$ is characterized by $\gamma_i$ and $\Gamma_i$. However, we set either the exponential or Gaussian decay parameter to $0$ to increase the stability of the fitting routines, and still find good agreement with the data. The QEC-DFS [[10,1,4]] decay is modeled as $F_q(t)=1/2+(1/2-\epsilon_{q,s})(1-2\epsilon_{q,m})^{t/\tau}$ where $\tau=2.89$~s. For comparisons, we note $F_q(t)\approx1/2+(1/2-\epsilon_{q,s})\text{exp}[-2\epsilon_{q,m}t/\tau]$ and report $2\epsilon_{q,m}\tau^{-1}$ in the table. We also note $\epsilon_{q,m}$ combines the idling and syndrome extraction error and leave it to future studies to characterize them separately. For all quantities, the index $i=\{p,d,q\}$ represents physical, [[2,1,1]], and [[10,1,4]] qubits respectively. $\epsilon_{i,s}$ are SPAM errors and $\epsilon_{q,m}$ is the logical error of the DFS-QEC qubit after a single idle period and QEC cycle. These quantities' definitions are further clarified in the Supplementary Material. The entries marked ``N/A" were not probed experimentally.}
\label{Table:FittingParameters}
\end{table}

\subsection*{Discussion}

\begin{figure}[ht]
 \centering
  \includegraphics[width=.7
  \linewidth]{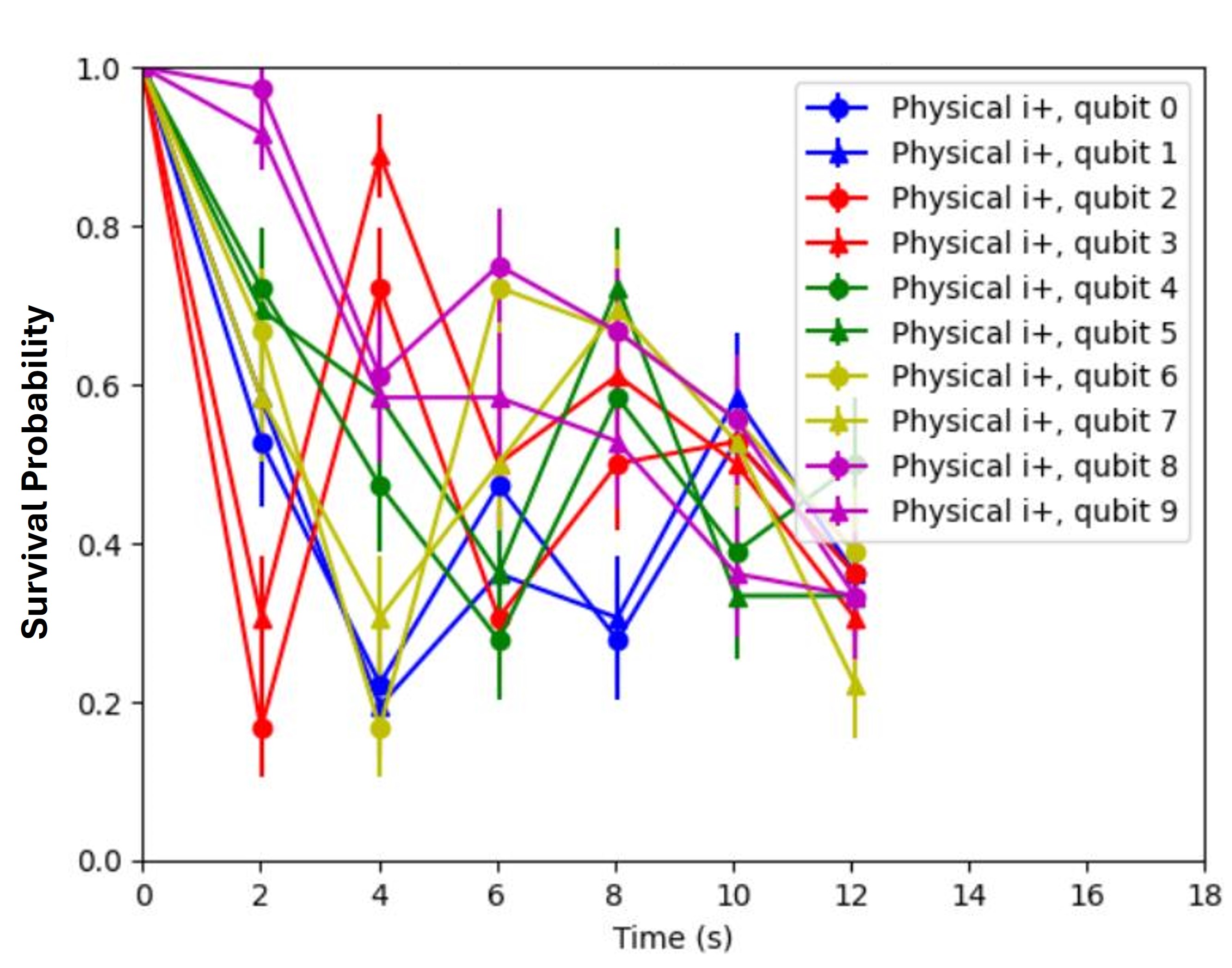}
\caption{\textbf{The survival probability for 10 qubits placed in the 5 gating regions.} The ions that are co-located are depicted using the same color.}
\label{fig:IndividualQubits}
\end{figure}

In this work, we've demonstrated the power of using DFS codes in conjunction with QEC. To demonstrate a gold standard logical memory~\cite{Xu2018}, an encoded memory should be superior for \textit{all} storage times, and showing this for short idling times practically requires small logical SPAM errors.  Previous experiments at Quantinuum~\cite{RyanAnderson2021,RyanAnderson2022} have demonstrated logical SPAM errors of $<10^{-3}$, and future work will aim to merge these techniques. Nonetheless, it is remarkable that our experiments show over an order-of-magnitude improvement in the qubit lifetime using ten data qubits for a logical qubit, whereas similar experiments with the surface code required forty-nine data qubits for a logical qubit to achieve a lifetime improvement of 2-3x~\cite{Acharya2024}. More generally, codes like the surface code will not extend the memory time through increased distances in an arbitrary manner without rapid QEC. Of course, \textit{with} rapid QEC, an increasing distance will result in a longer lived logical qubit. However, if we are interested in minimizing the resources needed to preserve a logical qubit, (e.g. laser pulses, ancilla qubits, syndrome decoding bandwidth), concatenation can offer favorable resource requirements over increased distances. For example, if we repeated our experiments with surface codes of increasing distance, no matter what the code distance is, approximately half the qubits would acquire a phase error after $\sim2$ s, thus violating the threshold criteria. Increasing the distance should be thought of as decreasing the logical error within a limited idle time window that respects the threshold criteria, whereas using a lower-level DFS encoding should be thought of as starting further below the threshold error and thus increasing the size of the idle time window that respects the threshold criteria. An optimal logical memory design then consists of allocating resources to both concatenation and distance scaling. Indeed, one can imagine designing individual concatenation layers to optimally tackle the dominant error eminating from the preceding layer. Machine time-constraints prevented us from probing the $[[5,1,3]]$ code, but we expect this code to be overwhelmed by the dominant noise-source, since it only corrects a single $Z$ error ($Z$-noise on 20\% of its qubits), and coherent dephasing causes Z-errors on a significant percentage of qubits by 2 seconds, as seen in Figs. \ref{fig:+StateFidelity}, \ref{fig:-StateFidelity}, \ref{fig:+iStateFidelity}, \ref{fig:-iStateFidelity}, and~\ref{fig:IndividualQubits}. Perhaps a reasonable proxy for the $[[5,1,3]]$ code performance compared to the physical level would be the increased performance of [[10,1,4]] over the [[2,1,1]] code, which is $\sim$2-3x.

Another interesting study would be to optimize the idle time before a QEC cycle is implemented, something we did not attempt. As discussed in Ref.~\cite{Xu2018}, it is not always advantageous to error correct as rapidly as possible since syndrome extraction injects noise. Without QEC, the logical error probability grows non-linearly in time, increasing slowly for short idle times and more rapidly for longer times. Therefore, there exists an optimal idle time to implement QEC such that resetting the non-linearity is worth the error from the QEC cycle, a tradeoff that will depend on the code and the available number of qubits. Due to the limited number of qubits in H1, we used flags to implement fault-tolerance in our QEC cycle, resulting in high-depth circuits and more memory error during the QEC cycle. Larger machines will have lower-depth options for QEC, including DFS CSS codes and Steane~\cite{Steane1997,Postler2024} or Knill-style~\cite{Knill2004,RyanAnderson2024} error correction.

There are many opportunities to improve the physical layer memory noise in future work, including dynamical decoupling~\cite{Viola1999,Biercuk2009}, randomized compiling~\cite{Wallman2016}, additional shielding, and more advanced calibrations. However, these improvements would also manifest at the logical level with possibly larger gains due to the non-linear relationship between physical and logical error rates, though eventually the logical error would be limited by gate noise during syndrome extraction. As alluded to earlier, improvements could also be made at the logical level, such as a transport compiler that attempts to keep DFS pairs co-located during re-sorting operations. There remains a large number of questions to answer, but this study clearly shows the power of DFS-QEC codes in quantum memories, paving a path for improved resource requirements in QCCD quantum computing architectures.

\clearpage




\clearpage 
\bibliography{science_template} 
\bibliographystyle{sciencemag}


\section*{Acknowledgments}
We thank the entire team at Quantinuum for their contributions toward making this research possible. We thank Yohei Matsuoka for helping to run these experiments.
\paragraph*{Funding:}
This work was internally funded by Quantinuum.
\paragraph*{Author contributions:}
C.G. and B.C. provided initial ideas and inspiration for the project. S.D. and D.H. conceived and designed the experiments and wrote the manuscript with contributions from N.L., G.V., J.G. and D.O.. S.D. developed the theory and quantum circuits. N.L., C.H., C.F., J.G., D.O., K.G., A. R. M. and G.V. were the experimentalists running the H1 device.
\paragraph*{Competing interests:}
There are no competing interests to declare.
\paragraph*{Data and materials availability:}
All data available in the manuscript, as well as the submitted QASM, is deposited at GitHub ~\cite{shivaldasu2025github}.


\begin{center}
\section*{Supplementary Materials for\\ \scititle}

	Shival~Dasu$^{\ast}$,
        Ben~Criger,
        Cameron ~Foltz,
        Justin~A.~Gerber,
        Chris~N.~Gilbreth,
        Kevin~Gilmore,
        Craig~A.~Holliman,
	Nathan~K.~Lysne,
        Alistair~R.~Milne,
        Daichi ~Okuno,
        Grahame~Vittorini,
        David~Hayes\\
	
	\small$^\ast$Corresponding author. Email: shival.dasu@quantinuum.com\and

\end{center}



\subsection*{Supplementary Text}


\subsubsection{Code definitions}
\begin{figure}[ht]
	\centering
	
    \includegraphics[width=0.6\textwidth]{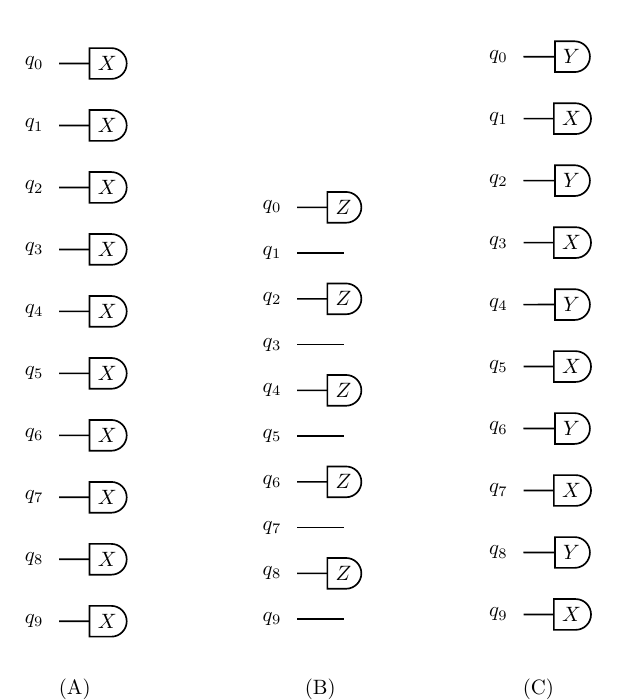}
	\caption{\textbf{Circuits for destructively measuring the logical operators of the [[10,1,4]] code.}
		Circuits for measuring \textbf{(A)} logical X, \textbf{(B)} logical Z, and  \textbf{(C)} logical Y.
	}
	\label{sup fig:Measure Out Circuits} 
\end{figure}

\begin{figure}[ht] 
	\centering

    \includegraphics[width=0.3\textwidth]{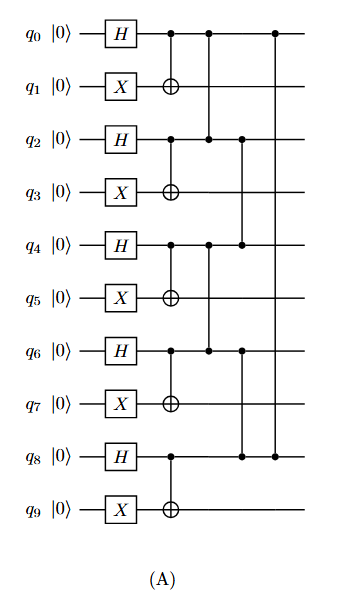}
    \includegraphics[width=0.565\textwidth]{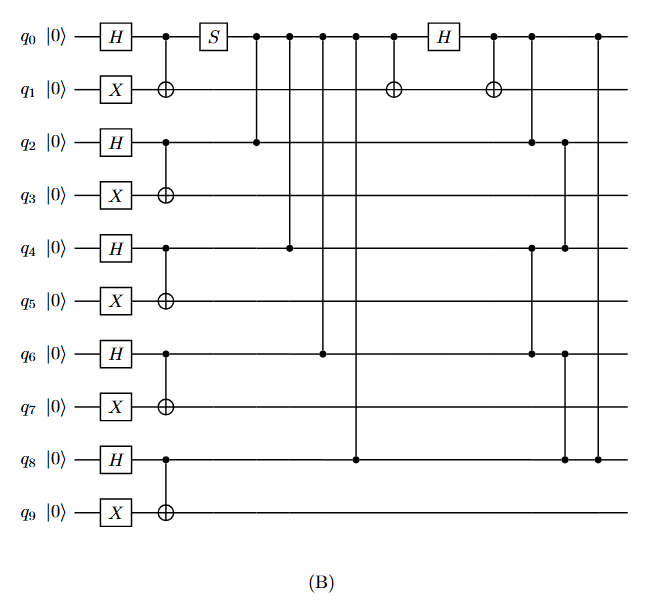}
	
	\caption{\textbf{Non-Fault Tolerant Initialization Circuits}
		(\textbf{A}) Initialization circuit for $\ket{-}_L$ in the [[10,1,4]] code (\textbf{B}) Initialization circuit for $\ket{i-}_L$. The circuit for $\ket{0}_L$ is identical to (B) except that there is no $S$ gate applied to qubit 0.
	}
	\label{sup fig:init circuits} 
\end{figure}

The [[10,1,4]] QEC code is obtained by concatenating the [[2,1,1]] DFS repetition code discussed in the introductory paragraph with stabilizers and logical operators given by $-Z_0 Z_1$ and $\{\bar{X} = X_0 X_1,\bar{Z} = Z_0\}$ respectively and the well-known [[5,1,3]] code~\cite{Laflamme96} generated by stabilizers $X_0 Z_1 Z_2 X_3$, $X_1 Z_2 Z_3 X_4$, $X_0 X_2 Z_3 Z_4$, and $Z_0 X_1 X_3 Z_4$ and possessing logical operators $\{\bar{X}=X_0X_1X_2X_3X_4,\bar{Z} = Z_0 Z_1 Z_2 Z_3 Z_4\}$. In the [[10,1,4]] code, the Pauli operators $X_i$ and $Z_i$ in the expressions for the $[[5,1,3]]$ stabilizers and logical operators become the logical operators of the $i$th DFS code. Therefore, the stabilizers for the [[10,1,4]] code are generated by $s_0 = \bar{X}_0\bar{Z}_1\bar{Z}_2\bar{X}_3$, $s_1 = \bar{X}_1\bar{Z}_2\bar{Z}_3\bar{X}_4$, 
$s_2 = \bar{X}_0\bar{X}_2\bar{Z}_3\bar{Z}_4$, and
$s_3 = \bar{Z}_0\bar{X}_1\bar{X}_3\bar{Z}_4$, 
which are equal to 
$X_0 X_1 Z_2 Z_4 X_6 X_7$,
$X_2 X_3 Z_4 Z_6 X_8 X_9$,
$X_0 X_1 X_4 X_5 Z_6 Z_8$, and 
$Z_0 X_2 X_3 X_6 X_7 Z_8$, as well as by the stabilizers from the repetition code on each pair of qubits, $r_0 = -Z_0Z_1, r_1 = -Z_2Z_3, r_2 = -Z_4Z_5, r_3 = -Z_6Z_7,$ and $r_4 = -Z_8Z_9$. The logical operators are $\bar{X}_{[[10,1,4]]} = \bar{X}_0\bar{X}_1\bar{X}_2\bar{X}_3\bar{X}_4 = X_0X_1X_2X_3X_4X_5X_6X_7X_8X_9$ and $\bar{Z}_{[[10,1,4]]} = \bar{Z}_0\bar{Z}_1\bar{Z}_2\bar{Z}_3\bar{Z}_4 = Z_0Z_2Z_4Z_6Z_8,$ where the $[[10,1,4]]$ subscripts are added to indicate that the bar refers to the logical operators for the $[[10,1,4]]$ code and not the $[[2,1,1]]$ code. Therefore the logical $Y$ operator for the $[[10,1,4]]$ code is $\bar{Y}_{[[10,1,4]]} = Y_0X_1Y_2X_3Y_4X_5Y_6X_7Y_8X_9.$ From the definitions of the logical operators, we see that the circuits for measuring $\bar{X}_{[[10,1,4]]}$, $\bar{Y}_{[[10,1,4]]}$, and $\bar{Z}_{[[10,1,4]]}$ are given in Fig. \ref{sup fig:Measure Out Circuits}. We also provide initialization circuits for the Pauli basis states of the [[10,1,4]] code in Fig. \ref{sup fig:init circuits}.\\
Finally, we remark on the evenness of the code distance. There are minimal weight logical operators for the [[5,1,3]] code are of the form $\bar{X} = Z_{i-1}X_iZ_{i+1}$ where the subscripts are taken mod 5. For the [[10,1,3]] code, these operators become $\bar{X} = Z_{2i- 2}X_{2i}X_{2i+1}Z_{2i +2}$ where the subscripts are mod 10. From this representation of $\bar{X}$, we can see why uncorrectable errors occur in this code due to its even distance. For example, if a $Z_{2i - 2}X_{2i}$ error occurred, then we are unable to differentiate between this error and $X_{2i+1}Z_{2i+2}$ because both errors have identical syndromes since they form a logical operator together. When decoding in real time, we guess randomly between these two possibilities if such errors occur. In post-processing, we also explore post-selecting on these uncorrectable weight 2 errors which are part of a weight 4 logical, as well as post-selecting on any syndrome where the lowest weight correction is at least 3, since protecting against such errors is beyond the ability of a distance 4 code in general.

\subsubsection{Syndrome Extraction}

We design our QEC gadget to be fault-tolerant to the distance of [[10,1,4]], i.e., to distance 4 with no post-selection on uncorrectable weight-2 errors. If $p$ is the probability of a single fault occurring, without post-selection, we are able to correct one fault, but there are some combinations of two faults which are not correctable in principle, so the error rate of an extended rectangle (QEC cycle followed by idling period, followed by QEC cycle) cannot be better than $O(p^2).$ 

For $d=4,$ the number of correctable errors is $t = \lfloor\frac{d-1}{2}\rfloor = 1$, and the fault-tolerance criteria in \cite{gottesman2009} are the same as for $d=3$ without post-selection. We designed our real-time decoder to guess randomly on uncorrectable weight-2 errors, but if we had designed our decoder to post-select on the corresponding syndromes instead the fault-tolerance criteria would be more stringent \cite{Prabhu_2024}. Nevertheless, we recorded all the syndrome information measured during the experiment and can explore post-selecting on such syndromes in postprocessing, as seen in the dashed lines in Figs. \ref{fig:Integrity}, \ref{fig:0StateFidelity}, \ref{fig:1StateFidelity},  \ref{fig:+StateFidelity}, \ref{fig:-StateFidelity}, and \ref{fig:+iStateFidelity}, and \ref{fig:-iStateFidelity}. 

Without post-selection, the fault-tolerance conditions are stated in \cite{Aliferis2005} as (a) if a QEC cycle has no faults, it takes any input to an output in the codespace, (b) if a QEC cycle has no faults, then it produces the correct output for an input with either no errors or a weight 1 error, and (c) if a single fault occurs during a QEC cycle, then it takes an input with no errors to an output with an error of at most weight 1. 

There is an additional fault-tolerance condition introduced in Ref.~\cite{Aliferis2005} stating that if one fault occurs in a QEC cycle, the QEC cycle must take an input to an output which is within a weight 1 Pauli error of some codeword (but the codeword doesn't have to be correct, so a logical error is allowed). Since we are not exploring higher-level concatenations of this code and are only interested in the failure rate of our protocol being $O(p^2)$, we can disregard this condition. As discussed in \cite{Aliferis2005}, conditions (a), (b), and (c) are sufficient to ensure that a logical error occurs in our experiments with probability $O(p^2)$, where $p$ is the probability of a single fault occurring.

Here, a ``fault" is either a single gate failure, single measurement error, or an error affecting a single qubit which occurs during idling. We do not include qubits leaking during two-qubit gates in our fault definition since this is small relative to the other errors. We leave designing protocols fault-tolerant to leakage to future work. Nevertheless, after each idling period, leakage detection is performed on all qubits via our leakage detection gadget~\cite{Moses2023}. Leaked qubits are reset, ensuring the subsequent QEC cycle converts the error to Pauli errors and send data qubits back into the codespace.

In this section, we show that our syndrome extraction circuits satisfy condition (c), and other conditions are addressed in the next section. 

\begin{figure}[ht]
	\centering
	
    \includegraphics[width=0.6\textwidth]{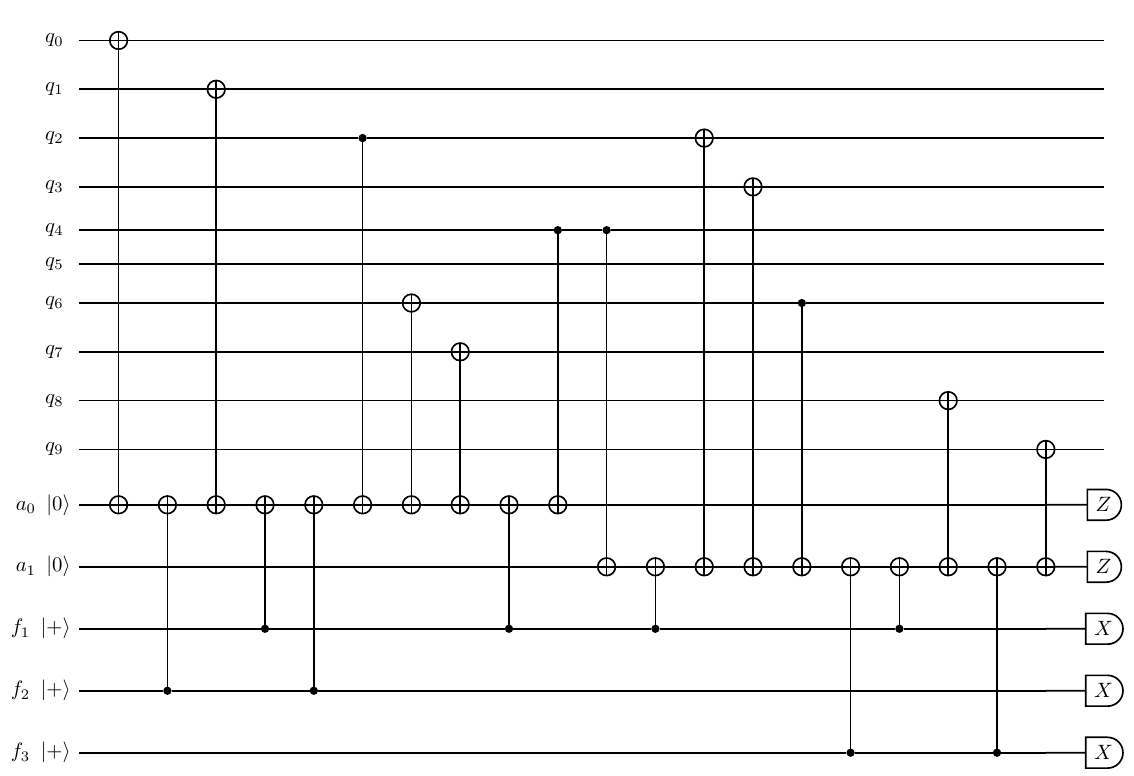}
	
	\caption{\textbf{Flagged Syndrome Extraction Circuit I}
		This circuit extracts with flags the first two of the syndromes of the encoded [[5,1,3]] code, $s_0 = \bar{X}_0\bar{Z}_1 \bar{Z}_2\bar{X}_3$, and $s_1 = \bar{X}_1\bar{Z}_2 \bar{Z}_3\bar{X}_4$.
	}
	\label{sup fig:fault-tolerant SE 1} 
\end{figure}

\begin{figure}[ht]
	\centering
	
    \includegraphics[width=0.6\textwidth]{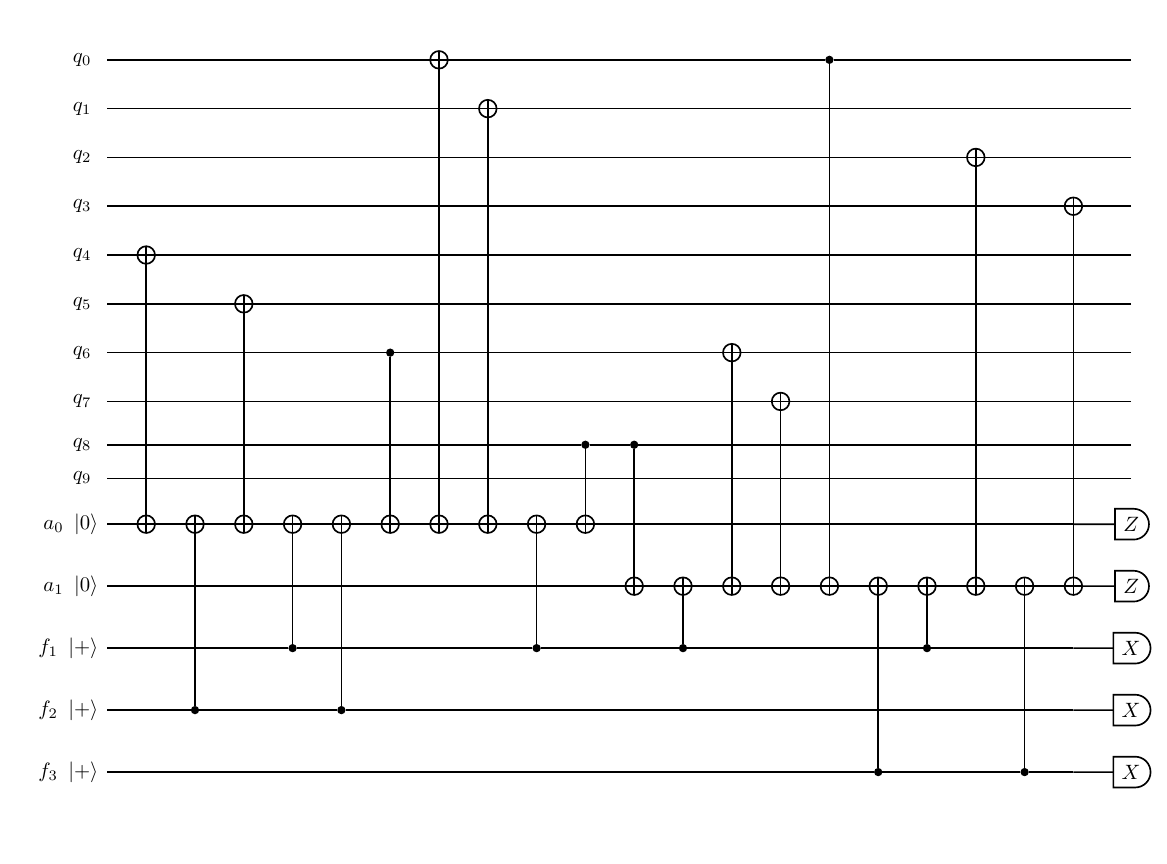}
	
	\caption{\textbf{Flagged Syndrome Extraction Circuit 2}
		This circuit extracts with flags the remaining two of the encoded [[5,1,3]] stabilizers, $s_1 = \bar{X}_0\bar{X}_2\bar{Z}_3 \bar{Z}_4$, and $s_1 = \bar{Z}_0 \bar{X}_1\bar{X}_3\bar{X}_4$.
	}
	\label{sup fig:fault-tolerant SE 2} 
\end{figure}

We design our circuits to satisfy condition (c) by preventing ``hook errors," which are errors of weight greater-than-one caused by single faults. We do this by adding flag registers $f_1, f_2,$ and $f_3$ in Figs.~\ref{sup fig:fault-tolerant SE 1} and \ref{sup fig:fault-tolerant SE 2}. If we model each gate failure as 2-qubit Pauli noise, hook errors come from $Z$ or $Y$ errors occurring on ancilla qubits $a_0$ or $a_1$, as it will propagate through CNOTs to Pauli errors on the code qubits. The flag measurements will detect these errors, and, if a hook error has occurred, we can correct it using the results of subsequent syndrome extraction, which we assume to be error-free since we are only trying to protect against a single fault. For more information, see \cite{Reichardt2018}.

\begin{figure}[ht] 
	\centering
	
    \includegraphics[width=0.9\textwidth]{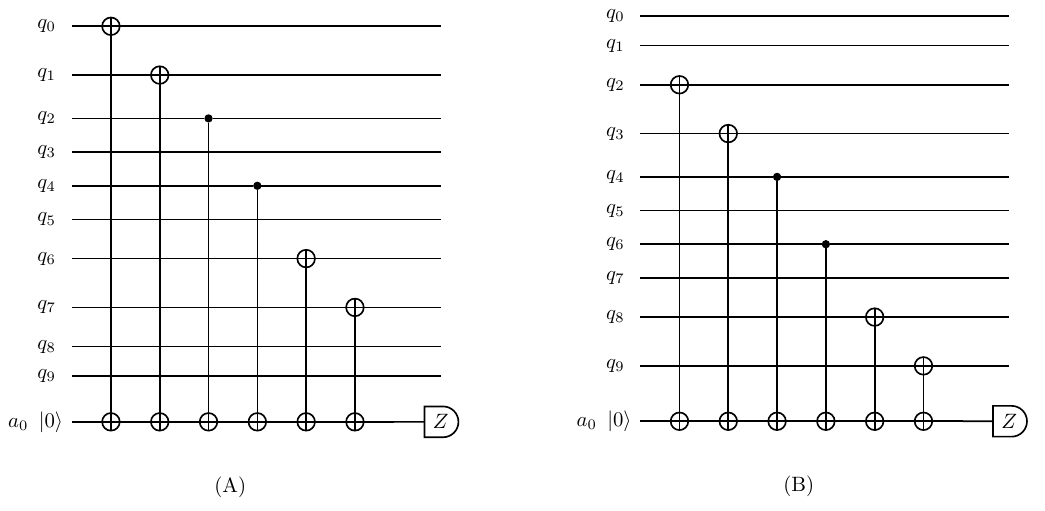}

	\caption{\textbf{Unflagged syndrome extraction circuits 1 and 2.}
		\textbf{(A)} Unflagged circuit for extracting the syndrome corresponding to stabilizer $s_0 = \bar{X_0}\bar{Z_1}\bar{Z_2}\bar{X_3} = X_0X_1Z_2Z_4X_6X_7$ of the [[10,1,4]] code. \textbf{(B)} Unflagged circuit for extracting the syndrome corresponding to $s_1 = \bar{X_1}\bar{Z_2}\bar{Z_3}\bar{X_4} = X_2X_3Z_4Z_6X_8X_9$.
	}
	\label{sup fig:unflaggedSE1} 
\end{figure}

\begin{figure} 
	\centering
    \includegraphics[width=0.9\textwidth]{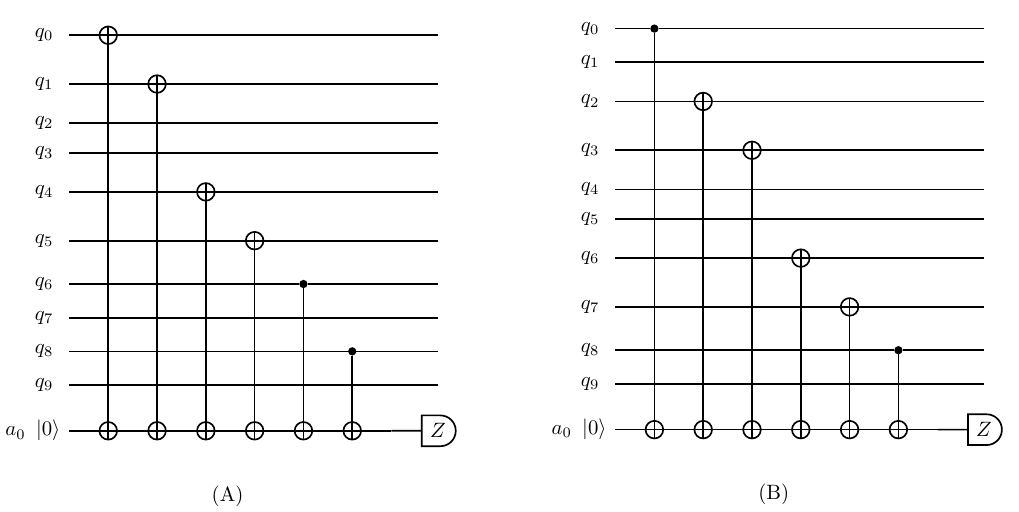}

	\caption{\textbf{Unflagged syndrome extraction circuits 3 and 4.}
		\textbf{(A)} Unflagged circuit for extracting the syndrome corresponding to stabilizer $s_2 = \bar{X_0}\bar{X_2}\bar{Z_3}\bar{Z_4} = X_0X_1 X_4X_5Z_6Z_8$ of the [[10,1,4]] code. \textbf{(B)} Unflagged circuit for extracting the syndrome corresponding to $s_3 = \bar{Z_0}\bar{X_1}\bar{X_3}\bar{X_4} = Z_0X_2X_3X_6X_7Z_8$.
	}
	\label{sup fig:unflaggedSE2} 
\end{figure}

\begin{figure}[ht] 
	\centering
	
    \includegraphics[width=0.6\textwidth]{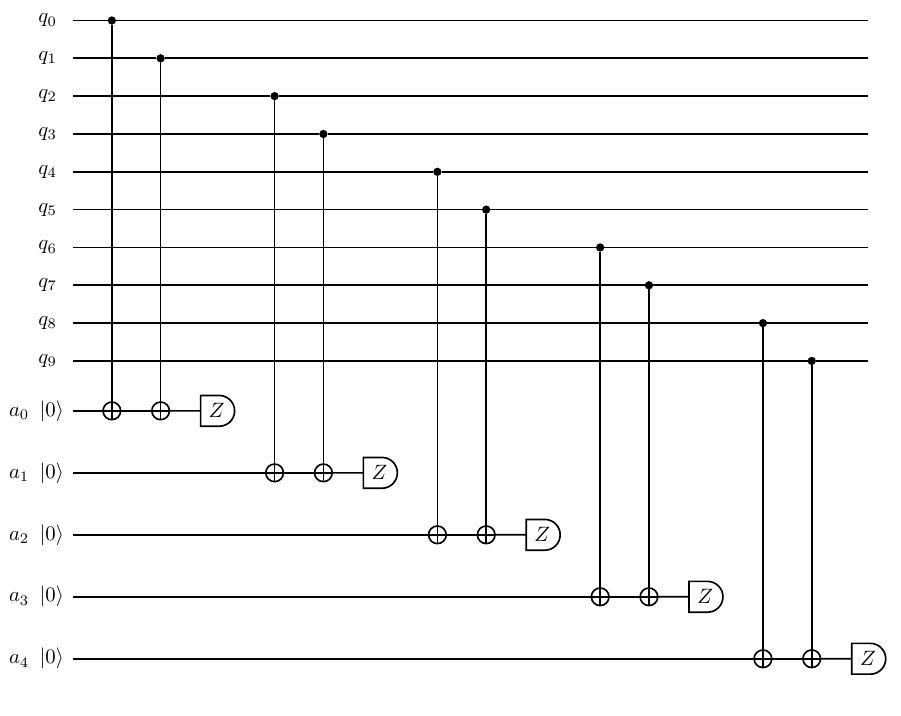}
	
	\caption{\textbf{Fault-tolerant circuit for extracting the DFS syndromes of the [[10,1,4]] code}
		This circuit fault-tolerantly extracts the syndromes for the five stabilizers of the [[10,1,4]] code inherited from the base code. These stabilizers are $r_0 = -Z_0Z_1$, $r_1 = -Z_2Z_3$, $r_2 = -Z_4Z_5$, $r_3 = -Z_6Z_7$ and $r_4 = -Z_7Z_8$. Note that since the circuit is measuring the positive products of $Z$s, $Z_iZ_{i+1}$, the syndromes extracted should all be $1$ in the absence of errors.
	}
	\label{sup fig:fault-tolerant SE DFS} 
\end{figure}

The DFS syndrome extraction circuits $r_0, r_1, r_2, r_3$ and $r_4$ in Fig. \ref{sup fig:fault-tolerant SE DFS} do not cause hook errors since either gate failing results in an error equivalent to a weight-1 error up to the DFS stabilizers. Therefore, our fault-tolerant QEC cycle consists of the following: After idling, leakage detection is performed on all qubits and any leaked qubit is reset so that subsequent syndrome extraction will send it back to the code space. Subsequently, the DFS syndromes are extracted via the circuit in Fig. \ref{sup fig:fault-tolerant SE DFS}. If a non-trivial syndrome is detected, all syndromes, including the DFS ones, are extracted using the unflagged circuits in Figs. \ref{sup fig:fault-tolerant SE DFS}, \ref{sup fig:unflaggedSE1}, and \ref{sup fig:unflaggedSE2}. If no non-trivial syndrome is detected, we proceed to the first flagged syndrome extraction circuit (Fig.~\ref{sup fig:fault-tolerant SE 1}) for $s_0$ and $s_1$. If any of these are non-trivial or if any of the flags $f_0, f_1$ or $f_2$ are non-zero, we extract all unflagged syndromes. If not, we proceed to the second flagged syndrome extraction circuit (Fig.~\ref{sup fig:fault-tolerant SE 2}) for $s_2$ and $s_3$. If either of the syndromes or any of the flags are non-trivial, we extract all the syndromes using the unflagged circuits. If there are non-trivial flag measurements, our decoder will use the unflagged syndrome information to correct hook errors via a lookup table to satisfy condition (c). We describe how our decoder works in the absence of hook errors in the next section.

Finally, we note that our QEC cycle makes extensive use of H1's mid-circuit measurement and reset (MCMR) capabilities. H1 uses twenty qubits, ten are used as data qubits in the [[10,1,4]] code and the other ten are ancillas that are reset multiple times during syndrome extraction. After every idling period, all ten ancillas are reset and then used to perform leakage detection on the data qubits. The ancillas are then measured and reset. Any data qubit found to have leaked is reset at this time. Then, five ancilla qubits are used to measure the DFS syndromes and then reset and the five additional ancillas are used to measure syndromes and flags in the first flagged syndrome extraction circuit (Fig. \ref{sup fig:fault-tolerant SE 1}). If there are no non-trivial measurements, the five ancillas used to measure the DFS syndromes are used for the second flagged syndrome extraction circuit (Fig. \ref{sup fig:fault-tolerant SE 2}). If there are any non-trivial syndromes, all ancilla qubits are reset and then eight of them are used to extract all the unflagged syndromes (Figs. \ref{sup fig:fault-tolerant SE DFS} \ref{sup fig:unflaggedSE1}, and \ref{sup fig:unflaggedSE2}). 

\subsubsection{Decoding}

\begin{figure}[ht]
 \centering
  \includegraphics[width=.7
  \linewidth]{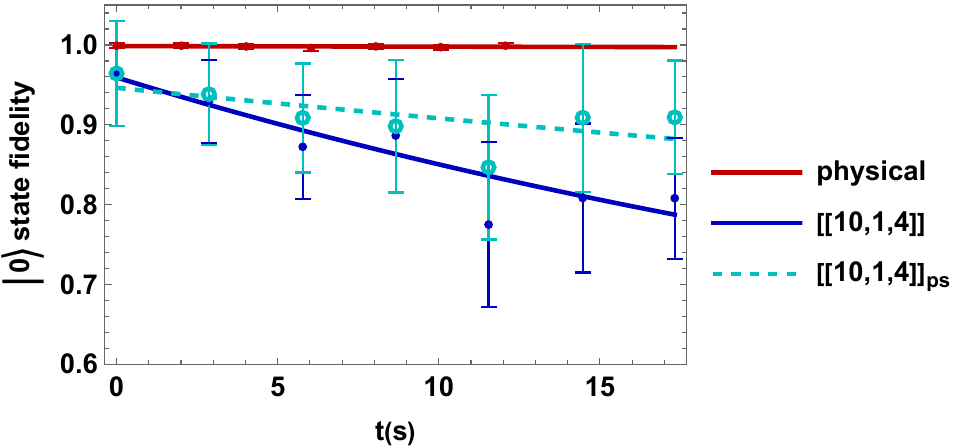}
\caption{\textbf{The $\ket{0}$ state fidelity for two different qubits investigated in this work.}}
\label{fig:0StateFidelity}
\end{figure}

\begin{figure}[ht]
 \centering
  \includegraphics[width=.7
  \linewidth]{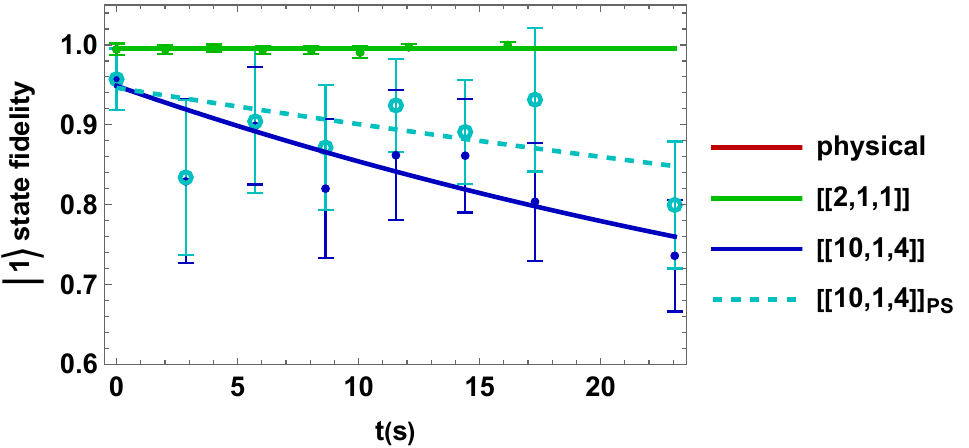}
\caption{\textbf{The $\ket{1}$ state fidelity for the three different qubits investigated in this work.} We note that for the physical qubits, the measurement does not distinguish between $\ket{1}$ and the leakage subspace of the $F=1$ manifold of the hyperfine qubit. Therefore, this data only represents an upper bound on the physical level survivability of $\ket{1}$. However, as discussed in the main text, the experiments probing the superposition states' memories are indeed sensitive to leakage errors.}
\label{fig:1StateFidelity}
\end{figure}

\begin{figure}[ht]
 \centering
  \includegraphics[width=.7
  \linewidth]{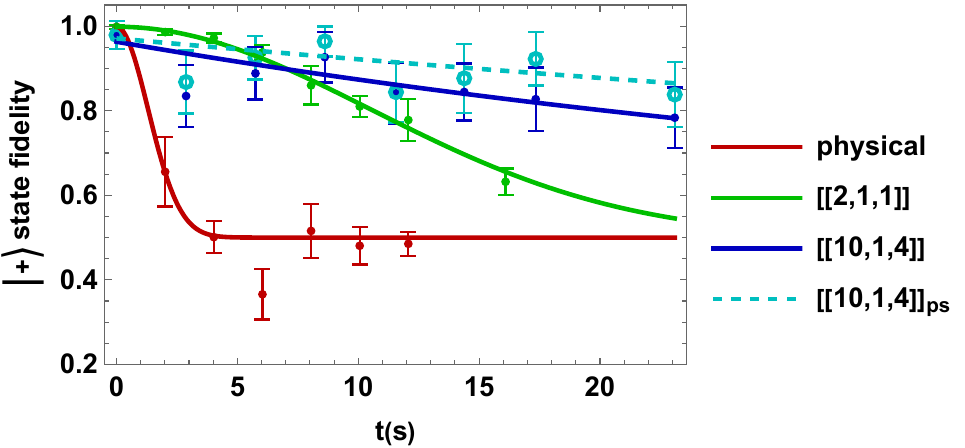}
\caption{\textbf{The $\ket{+}$ state fidelity for the three different qubits investigated in this work.} The physical and DFS qubit infidelities are dominated by coherent phase errors.}
\label{fig:+StateFidelity}
\end{figure}

\begin{figure}[ht]
 \centering
  \includegraphics[width=.7
  \linewidth]{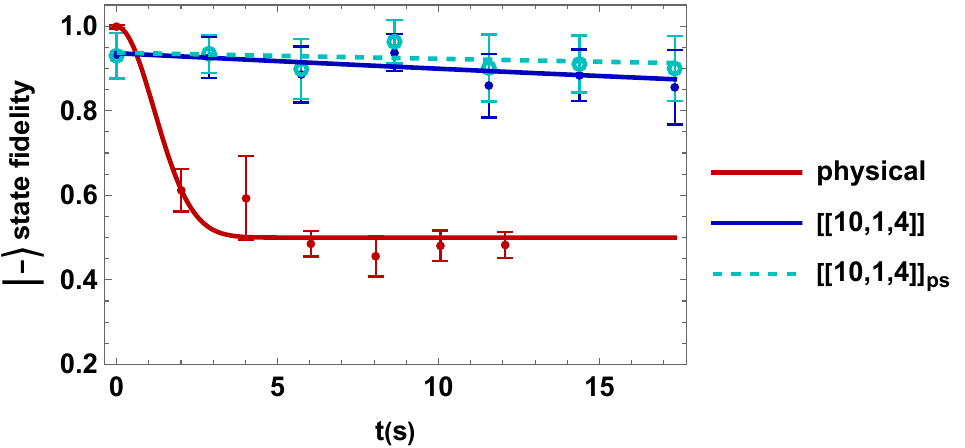}
\caption{\textbf{The $\ket{-}$ state fidelity for two different qubits investigated in this work.} The physical and DFS qubit infidelities are dominated by coherent dephasing errors.}
\label{fig:-StateFidelity}
\end{figure}

\begin{figure}[ht]
 \centering
  \includegraphics[width=.7
  \linewidth]{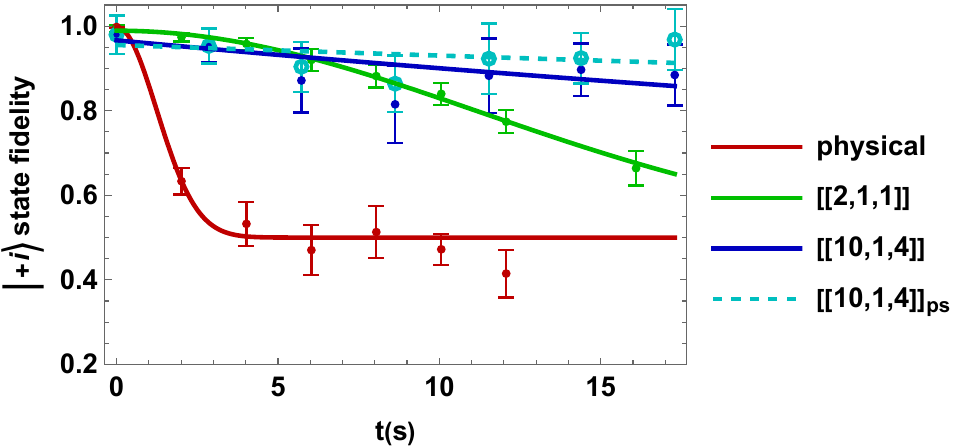}
\caption{\textbf{The $\ket{+i}$ state fidelity for the three different qubits investigated in this work.} The physical and DFS qubit infidelities are dominated by coherent dephasing errors.}
\label{fig:+iStateFidelity}
\end{figure}

\begin{figure}[ht]
 \centering
  \includegraphics[width=.7
  \linewidth]{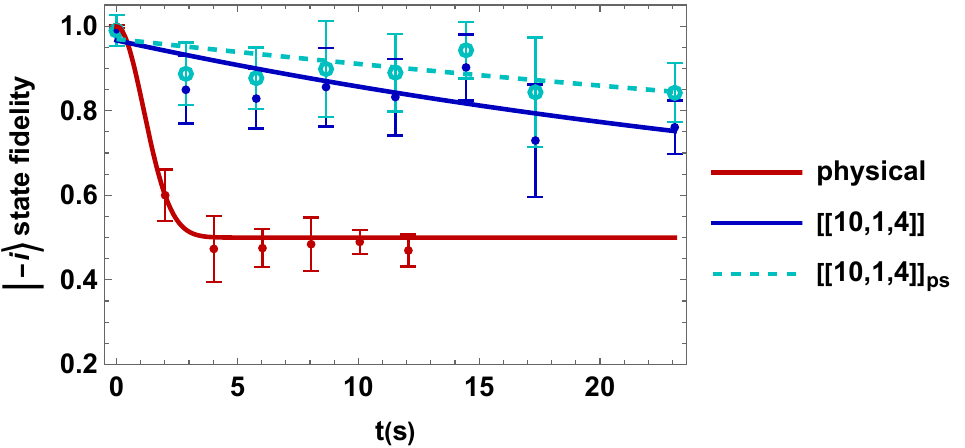}
\caption{\textbf{The $\ket{-i}$ state fidelity for two different qubits investigated in this work.} The physical and DFS qubit infidelities are dominated by coherent dephasing errors.}
\label{fig:-iStateFidelity}
\end{figure}

In the previous section, we discussed how our decoder works in the case of a single failure during the QEC cycle. We now turn to the case where the QEC cycle has no errors, but there are possibly errors on the input. We wish to correct any weight-1 error, but wish to use the fact that $Z$ errors are more likely than $X$ errors during idling, even taking into account the DFS protection against coherent dephasing. This can be seen in the survival plots of the $\ket{0}$ and $\ket{1}$ DFS states in Figs. \ref{fig:0StateFidelity} and \ref{fig:1StateFidelity}, where there is essentially no decay because these states are sensitive only to $X$ errors, but there is significant decay in the survival rates of other DFS states which are sensitive to Z noise, seen in Figs. \ref{fig:+StateFidelity}, \ref{fig:-StateFidelity}, \ref{fig:+iStateFidelity}, and \ref{fig:-iStateFidelity}. Rather than implementing a lookup table for each of the 512 unflagged syndromes, we use an algorithmic decoder which is run in real-time during each round of QEC.

\begin{figure}[ht] 
	\centering
    \includegraphics[width=0.3\textwidth]{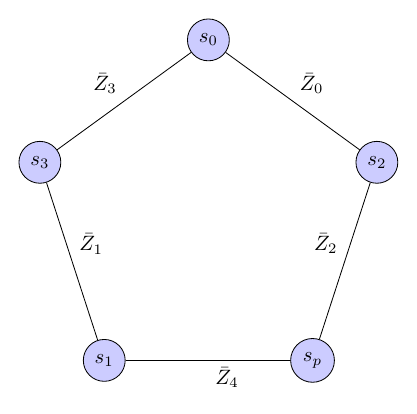}

	\caption{\textbf{Decoding Graph for $Z$ errors in the [[10,1,4]] code}
		As explained in the decoding section of the Supplementary Materials, there is one node per stabilizer $s_i$ of the [[10,1,4]] code. Each $\bar{Z}_i$ flips the signs of two of these stabilizers and an edge between those two stabilizers is drawn in the graph.
	}
	\label{sup fig:Decoding Graph} 
\end{figure}

We begin with the observation that $Z$ errors are graph-like for the syndromes of the [[5,1,3]] or [[10,1,4]] codes. This means that each $Z$ error flips at most two syndromes\cite{Higgott_2025}. If we introduce a redundant stabilizer $s_p = s_0s_1s_2s_3$, then every $Z$ error flips exactly two syndromes. For the [[10,1,4]] code, it suffices to consider the even-index $Z$ errors: $\bar{Z}_0 = Z_0, \bar{Z_1} = Z_2, \bar{Z}_2 = Z_4, \bar{Z}_3 = Z_6, \bar{Z}_4 = Z_8$. This is because the odd index $Z_i$ are equivalent to the even ones up to sign due to the DFS stabilizers, e.g., $Z_1 = -Z_0$ up to the stabilizer. We can form a graph where each syndrome is a node. For each $\bar{Z}_i$, we place a corresponding edge between the two syndromes that it flips. This graph is shown in Fig. \ref{sup fig:Decoding Graph}.

For any syndromes $s_0 = m_0, s_1 = m_1, s_2 = m_2, s_3 = m_3$, this graph can find the minimum weight $Z$ error compatible with $m_i$. We do this via 1-D minimum weight perfect matching on a cyclic graph, which is implemented as follows. First compute $s_p = m_0 + m_1 + m_2 + m_3$ mod 2 to ensure there are an even number of non-zero syndromes. To find a set of edges $S$ which is compatible with syndrome measurements $m_i$, we start at a non-zero syndrome and keep adding edges, traveling clockwise, until we reach the next non-zero syndrome. This pairs two of the non-zero syndromes with a path of edges $S_1.$ We repeat this with the next pair of non-zero syndromes and continue until we have paths linking them all. The set of edges $S$ is the union of the edges in all these paths. The two edge sets compatible with the non-zero syndromes are $S$ and $S^c$, the complement of $S$ in $E.$ The smaller of the two is the minimum set of $Z$ errors compatible with the syndromes.

As an example, suppose the only non-zero syndrome measured is $s_1.$ Then $s_p$ must also be non-zero. There are two paths between these syndromes in the decoding graph. Either we take $S = \{\bar{Z}_4$ \} or $S^c = \{\bar{Z}_0, \bar{Z}_1, \bar{Z}_2, \bar{Z}_3\}$. Clearly $\bar{Z}_4 = Z_8$ is the shorter path and is the minimum weight $Z$ error compatible with the measured syndromes. To arrive at the full decoder for [[10,1,4]], we must also decode $X$ errors. If a single $X$ error occurs, it will flip one of the DFS stabilizers. Conversely, if the sign of $r_i = -Z_{2i}Z_{2i+1}$ flips, we know either an $X_{2i}$ or $X_{2i+1}$ error occurred. If $m$ of the $r_i$ are non-zero, there are $2^m$ minimal combinations of $X$ errors compatible with those syndromes. For each set of $X$ errors, we solve a 1-D matching problem to find the minimum set of $Z$ errors compatible with the candidate $X$ errors and the syndromes $s_i$. Of these candidates, we pick one of the $X$ and $Z$ combinations where the set of $Z$ errors is the smallest.

We illustrate via the following example where an error occurred before syndrome extraction. We extract the flagged syndromes and find non-trivial syndromes, but no flags are flipped. We then extract the unflagged syndromes and find $s_0 = 1, s_1 = 0, s_2 = 0, s_3 = 1, s_p = 0$ and $r_0 = 1, r_1 = 0, r_2 =0, r_3=0, r_4 =0.$ From the signs of the $r_i$, the minimum weight $X$ error that occurred is either $X_0$ or $X_1$. If $X_0$ occurred, the $Z$ component of the error must flip $s_0$ and $s_p$ since $X_0$ flips $s_3$ and $s_p.$ Consulting our decoding graph, the minimum weight $Z$ error which does this is $\bar{Z}_0\bar{Z}_2 = Z_0Z_4$ Therefore, the first candidate is $X_0Z_0Z_4.$ If $X_1$ occurred, the $Z$ component of the error must flip $s_0$ and $s_3$ since $X_1$ does not flip any of the $s_i.$ The minimum Z error compatible with this is $\bar{Z}_3 = Z_6.$ Therefore, we decide between $X_0Z_0Z_4$ and $X_1Z_6$ and choose $X_1Z_6$ since it has the smaller number of $Z$ errors. This decoding strategy will correct any weight-1 error and send inputs back to the code space in the absence of any errors during syndrome extraction. Therefore, our QEC cycle satisfies conditions (a) and (b), completing the demonstration of the fault-tolerance of our QEC cycle.

This decoder will also correct any weight 2 $Z$ error, which is why we included the extra flags $f_2$ and $f_3$ in the syndrome extraction circuits. Otherwise, the second $X$-controlled-$X$ gate in Fig. \ref{sup fig:fault-tolerant SE 1} for instance could introduce a hook error of the form $\bar{X_0}.$ We could modify our decoder to handle this case in the absence of additional flags and still preserve the $O(p^2)$ failure property, but we would lose the ability to differentiate between a $\bar{Z}_1\bar{Z}_4$ error and a $\bar{X}_0$, making the code less robust to $Z$ noise. 
We cannot flag the second $X$-controlled-$X$ gate with $f_1$ because one possible hook error from this gate $X_0Y_1$ has the same syndrome as a hook error $Z_4Z_7$ from the fourth $X$-controlled-$X$ gate failing, but these two hook errors have different logical actions on the codespace, so we cannot correct both by knowing that flag $f_1$ flipped. Interestingly, our strategy is not possible for the $[[5,1,3]]$ code, as fault-tolerance forces us to correct the $s_0 =0, s_1=0, s_2=0, s_3=1$ syndrome as though an $X_0$ error occurred in that case, however more likely a $Z_1Z_4$ memory error may be.

Due to the fact that there may be multiple minimal $Z$-weight candidates, there are certain weight 2 errors for which this decoder guesses the outcome randomly. This is a consequence of the fact that the code has even distance. These are weight-2 errors of the form $X_iZ_l$ which form a weight-4 logical operator together with Pauli operators of the form $X_jZ_p$, where $X_iX_j$ is a logical operator of some DFS pair. In post-processing, we also investigate performance when post-selecting on these weight-2 errors instead of guessing, as well as on weight-2 $Z$ errors, as these can also be completed to a weight-4 logical. We also post-select on any error of weight higher than 2. The post-selected results are represented by the dashed lines in Figs. \ref{fig:Integrity}, \ref{fig:0StateFidelity}, \ref{fig:1StateFidelity},  \ref{fig:+StateFidelity}, \ref{fig:-StateFidelity}, \ref{fig:+iStateFidelity}, and \ref{fig:-iStateFidelity}.

Finally, we would like to point out that, while our decoder adds essentially no time overhead for the [[10,1,4]] code (real-time metrics from the device indicated each call to the decoder took 75 $\mu$s), this strategy grows exponentially in the number of DFS pairs and will be less tractable for larger code blocks with many DFS pairs. Nevertheless, a similar strategy with polynomial scaling may be employed in the future by decoding using belief propagation for concatenated codes~\cite{Poulin_2006}.

\subsubsection{Decay models}
Here we discuss the fitting models for the three different types of experiments. For the physical layer quantum memory, we assume the error to be dominated by $\hat{Z}$ noise. Without loss of generality, we take the $\ket{+}$ state as our example. After initialization, the $i^{th}$ qubit has state $\rho^{(i)}(0)=\frac{1}{2}(\ket{0}\bra{0}+\ket{1}\bra{1}+\ket{0}\bra{1}+\ket{1}\bra{0})$. After time $t$, the coherence terms accumulate an uncontrolled phase that we assume stems from two frequencies, $\delta\omega_s^{(i)}$ and $\delta\omega_f^{(i)}$, respectively representing slowly and fastly varying qubit splitting frequencies. If we ensemble averaging over many experiments and assume the random phase acquired by $\delta\omega_f^{(i)}$ to be Gaussian distributed and the frequency to have a white noise spectrum~\cite{Young2012}, then an exponential decay results whose time scale $\gamma$ is related to the auto-correlation of $\delta\omega_f^{(i)}$, giving,
\begin{equation}
\rho^{(i)}(t)=\frac{1}{2}(\ket{0}\bra{0}+\ket{1}\bra{1}+e^{-\gamma t}(e^{-i\delta\omega_s^{(i)}t}\ket{0}\bra{1}+e^{i\delta\omega_s^{(i)}t}\ket{1}\bra{0})).
\end{equation}

While the single qubit density matrix exhibits an exponential decay modulated by sinusoidal function, the quasi-static offset frequency is not constant across the entire device, reflecting a spatial inhomogeneity in the quasi-static uncompensated magnetic field. Additionally, these experiments were carried out over several days, during which the ambient magnetic field was recalibrated many times, likely resulting in different spatial inhomogeneities in the quasi-static error field. Due to the different offset frequencies in time and space, we ensemble average over a normal distribution of $\delta\omega_s^{(i)}$ values characterized by width $\Gamma$, and use the identity 
\begin{equation}
\frac{1}{\sqrt{2\pi\Gamma}}\int d\delta\omega_s^{(i)}\text{exp}\left[-\frac{1}{2}\left(\frac{\delta\omega_s^{(i)}}{\Gamma}\right)^2\right]e^{-i\delta\omega_s^{(i)}t}=\text{exp}\left[-\frac{1}{2}(\Gamma t)^2\right],
\end{equation}
to arrive at 
\begin{equation}
\rho(t)=\langle\rho^{(i)}(t)\rangle_i=\frac{1}{2}\left(\ket{0}\bra{0}+\ket{1}\bra{1}+e^{-\gamma t-\frac{1}{2}(\Gamma t)^2}\left(\ket{0}\bra{1}+\ket{1}\bra{0}\right)\right).
\end{equation}

Therefore, in our physical layer memory experiments we use a model that includes both an exponential and a Gaussian decay,

\begin{equation}
F_p(t)=\frac{1}{2}+\left(\frac{1}{2}-\epsilon_{p,s}\right)e^{-\gamma_p t-\frac{1}{2}(\Gamma_p t)^2}.
\end{equation}

Likewise, the decay model for the DFS code is,

\begin{equation}
F_{d}(t)=\frac{1}{2}+\left(\frac{1}{2}-\epsilon_{d,s}\right)e^{-\gamma_{d} t-\frac{1}{2}(\Gamma_{d} t)^2}.
\end{equation}

For the DFS-QEC code experiments, we follow Ref.\cite{RyanAnderson2021} and use an exponential decay in the number of QEC $+$ wait cycles, $n$,

\begin{equation}
F_{q}(n)=\frac{1}{2}+\left(\frac{1}{2}-\epsilon_{q,s}\right)(1-2\epsilon_{q,m})^{n}.
\end{equation}

In these expressions, $\epsilon_{i,s}$ is the SPAM error for the $i=\{p,d,q\}$ qubit and $\epsilon_{p,m}$ is the QEC-DFS error accrued in a single 2s idle plus QEC cycle operation. We fit these expressions to the experimental data and report the fitting parameters in Table~\ref{Table:FittingParameters}. When comparing the DFS-QEC model with the other two models, we plot in terms of the wall-clock time $t$, and so we substitute $n=t/T$ where $T$ is the time required for a QEC cycle circuit to be performed plus an idle time, which is equal to 2.89s on average with a standard deviation of $\leq50$ms.

\subsubsection{Additional data plots}

\newpage


\renewcommand{\thefigure}{S\arabic{figure}}
\renewcommand{\thetable}{S\arabic{table}}
\renewcommand{\theequation}{S\arabic{equation}}
\renewcommand{\thepage}{S\arabic{page}}
\setcounter{figure}{0}
\setcounter{table}{0}
\setcounter{equation}{0}
\setcounter{page}{1} 




\clearpage 


\end{document}